\documentclass[aps,prl,reprint,groupedaddress, showpacs]{revtex4-1}

\usepackage{graphicx}
\usepackage{epstopdf}
\usepackage{amssymb}
\usepackage{amsmath}
\usepackage{bm}
\usepackage{braket}

\begin{document}

\title{Universal space-time scaling symmetry in the dynamics of bosons across a quantum phase transition} 

\author{Logan W. Clark, Lei Feng, Cheng Chin}
\affiliation{James Franck Institute, Enrico Fermi Institute and Department of Physics, University of Chicago, Chicago, IL 60637, USA}

\date{\today}

\begin{abstract}
The dynamics of many-body systems spanning condensed matter, cosmology, and beyond is hypothesized to be universal when the systems cross continuous phase transitions. 
The universal dynamics is expected to satisfy a scaling symmetry of space and time with the crossing rate, inspired by the Kibble-Zurek mechanism.
We test this symmetry based on Bose condensates in a shaken optical lattice. Shaking the lattice drives condensates across an effectively ferromagnetic quantum phase transition. 
After crossing the critical point, the condensates manifest delayed growth of spin fluctuations and develop anti-ferromagnetic spatial correlations resulting from sub-Poisson generation of topological defects. The characteristic times and lengths scale as power-laws of the crossing rate, yielding the temporal exponent 0.50(2) and the spatial exponent 0.26(2), consistent with theory. Furthermore, the fluctuations and correlations are invariant in scaled space-time coordinates, in support of the scaling symmetry of quantum critical dynamics. 
\end{abstract}

\maketitle
Critical phenomena near a continuous phase transition reveal fascinating connections between seemingly disparate systems that can be described via the same universal principles. Examples exist in superfluid helium \cite{Zurek1985}, liquid crystals \cite{Chuang1991}, biological cell membranes \cite{Veatch2007}, the early universe \cite{Kibble1976}, and cold atoms \cite{Zhou2010,Polkovnikov2011}. An important prediction is the power-law scaling of the topological defect density with the rate of crossing a critical point, as first discussed by T.~Kibble in cosmology \cite{Kibble1976} and extended by W.~Zurek in the context of condensed matter \cite{Zurek1985}. Their theory, known as the Kibble-Zurek mechanism, has been the subject of intense experimental study that has largely supported the scaling laws \cite{delCampo2014}. Recent theoretical works further propose the universality hypothesis that the collective dynamics across a critical point should be invariant in the space and time coordinates that scale with the Kibble-Zurek power-law \cite{Kolodrubetz2012,Chandran2012,Francuz2016}.

Atomic quantum gases provide a clean, well-characterized, and controlled platform for studying critical dynamics \cite{Bloch2008,Dziarmaga2010,Polkovnikov2011}. They have enabled experiments on the formation of topological defects across the Bose-Einstein condensation transition \cite{Weiler2008,Lamporesi2013,Corman2014,Navon2015} as well as critical dynamics across quantum phase transitions \cite{Sadler2006,Baumann2011,Chen2011,Nicklas2015,Braun2015,Meldgin2016,Anquez2015}. Recent experiments using cold atoms in shaken optical lattices \cite{Gemelke2005,Lignier2007,Struck2012} have provided a new vehicle for exploring phase transitions in spin models \cite{Struck2011,Parker2013,Jotzu2014}.

In this report we study the critical dynamics of Bose condensates in a shaken optical lattice crossing an effectively ferromagnetic quantum phase transition. 
The transition occurs when we ramp the shaking amplitude across a critical value, causing the atomic population to bifurcate into two pseudo-spinor ground states \cite{Parker2013}. 
We measure the growth of spin fluctuations and the spatial spin correlations for ramping rates varied over two orders of magnitude. Beyond the critical point we observe delayed development of spin domains with anti-ferromagnetic correlations, a distinctive feature of the non-equilibrium dynamics. The times and lengths characterizing the critical dynamics agree excellently with the scaling predicted by the Kibble-Zurek mechanism. We further observe that the measured fluctuations and correlations collapse onto single curves in scaled space and time coordinates, supporting the universality hypothesis. 

\begin{figure*}
\includegraphics[width=130mm]{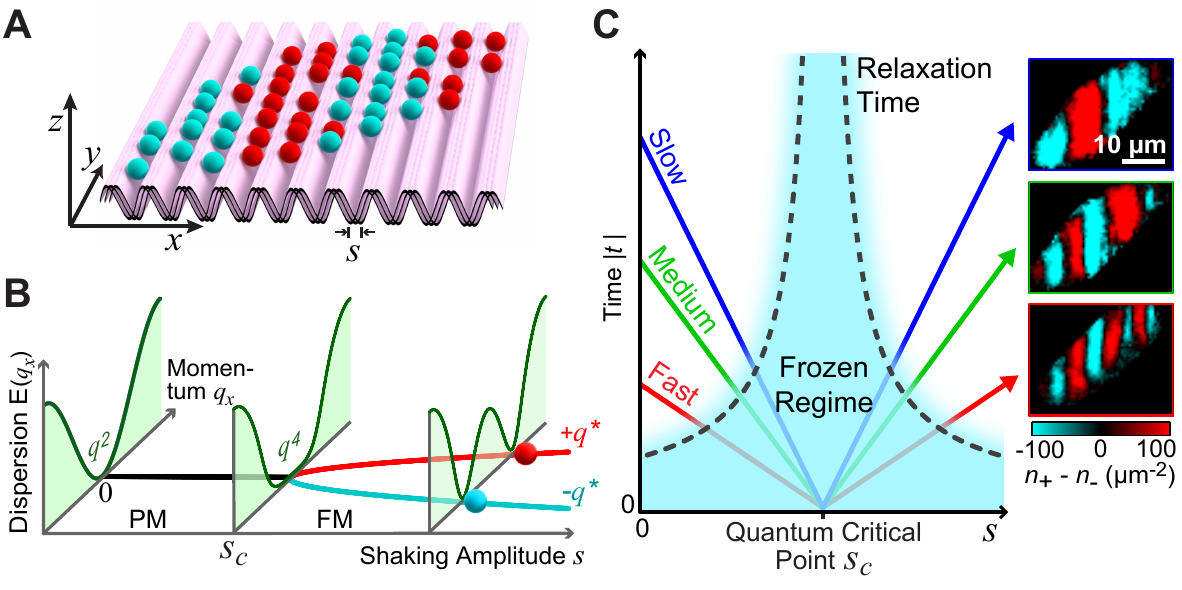}

\caption{\label{Fig1} Ferromagnetic quantum phase transition of bosons in a shaken optical lattice. (\textbf{A}) A BEC of cesium atoms (spheres) in a 1D optical lattice (pink surface) shaking with peak-to-peak amplitude $s$ can form ferromagnetic domains (blue and red regions). (\textbf{B}) The transition occurs when the dispersion evolves from quadratic for $s<s_\mathrm{c}$ (paramagnetic (PM) phase), through quartic at the quantum critical point $s=s_\mathrm{c}$, to a double-well for $s>s_\mathrm{c}$ (ferromagnetic (FM) phase) with two minima at $q_x=\pm{q}^*$ \cite{Parker2013}. (\textbf{C}) Evolution of the condensate crossing the phase transition becomes diabatic in the frozen regime (cyan) when the time $t$ remaining to reach the critical point is less than the relaxation time. Faster ramps cause freezing farther from the critical point, limiting the system to smaller domains. Sample domain images are shown for slow, medium, and fast ramps.}
\end{figure*}

Our experiments utilize Bose-Einstein condensates (BECs) of cesium atoms. We optically confine the condensates with trap frequencies of $(\omega_{x'}, \omega_{y'}, \omega_z) = 2\pi \times (12,~30,~70)$~Hz, where the long ($x'$) and short ($y'$) axes are oriented at $45^o$ with respect to the $x$ and $y$ coordinates (Fig.~\ref{Fig1}A). The tight confinement along the vertical $z$-axis suppresses non-trivial dynamics in that direction, which is also the optical axis of our imaging system. We adiabatically load the condensates into a one-dimensional (1D) optical lattice \cite{Bloch2008} along the $x$-axis with a lattice spacing of $532~$nm and a depth of 8.86~$E_\mathrm{R}$, where $E_\mathrm{R}=h~\times~1.33$~kHz is the recoil energy and $h$ is Planck's constant.

To induce the ferromagnetic quantum phase transition, we modulate the phase of the lattice beam to periodically translate the lattice potential by $\Delta{x}(t)=(s/2)\mathrm{sin}(\omega{t})$, where $s$ is the shaking amplitude and the modulation frequency $\omega$ is tuned to mix the ground and first excited lattice bands \cite{Parker2013,Zheng2014}. The hybridized ground band dispersion $\epsilon$ can be modelled for small quasimomentum $\mathbf{q}=(q_x,~q_y,~q_z)$ by 

\begin{equation}
\label{eqn:KE}
\epsilon(\mathbf{q}; s) = \alpha(s)q_x^2 + \beta(s)q_x^4 + \frac{q_y^2+q_z^2}{2m},
\end{equation}

\noindent where $m$ is the atomic mass, and the coefficients of its quadratic ($\alpha$) and quartic ($\beta$) terms depend on the shaking amplitude (Fig.~\ref{Fig1}B).
For shaking amplitudes below the critical value the coefficient $\alpha$ is positive and the BEC occupies the lone ground state at momentum $\mathbf{q}=0$. The quantum phase transition occurs when the quadratic term crosses zero at $s=s_c$, where $\alpha=0$ and $\beta>0$. Stronger shaking converts the dispersion into a double-well with $\alpha<0$, yielding two degenerate ground states with $q_x=\pm{q^*}$. Repulsively-interacting bosons with this double-well dispersion are effectively ferromagnetic, leaving two degenerate many-body ground states with all atoms either pseudo-spin up ($q_x=q^*$) or down ($q_x=-q^*$). Notably, transitioning to one of these two ground states requires the system to spontaneously break the symmetry of its Hamiltonian. Describing the dynamics across the critical point presents a major challenge due to the divergence of the correlation length and relaxation time (critical slowing). 

\begin{figure*}
\includegraphics[width=135mm]{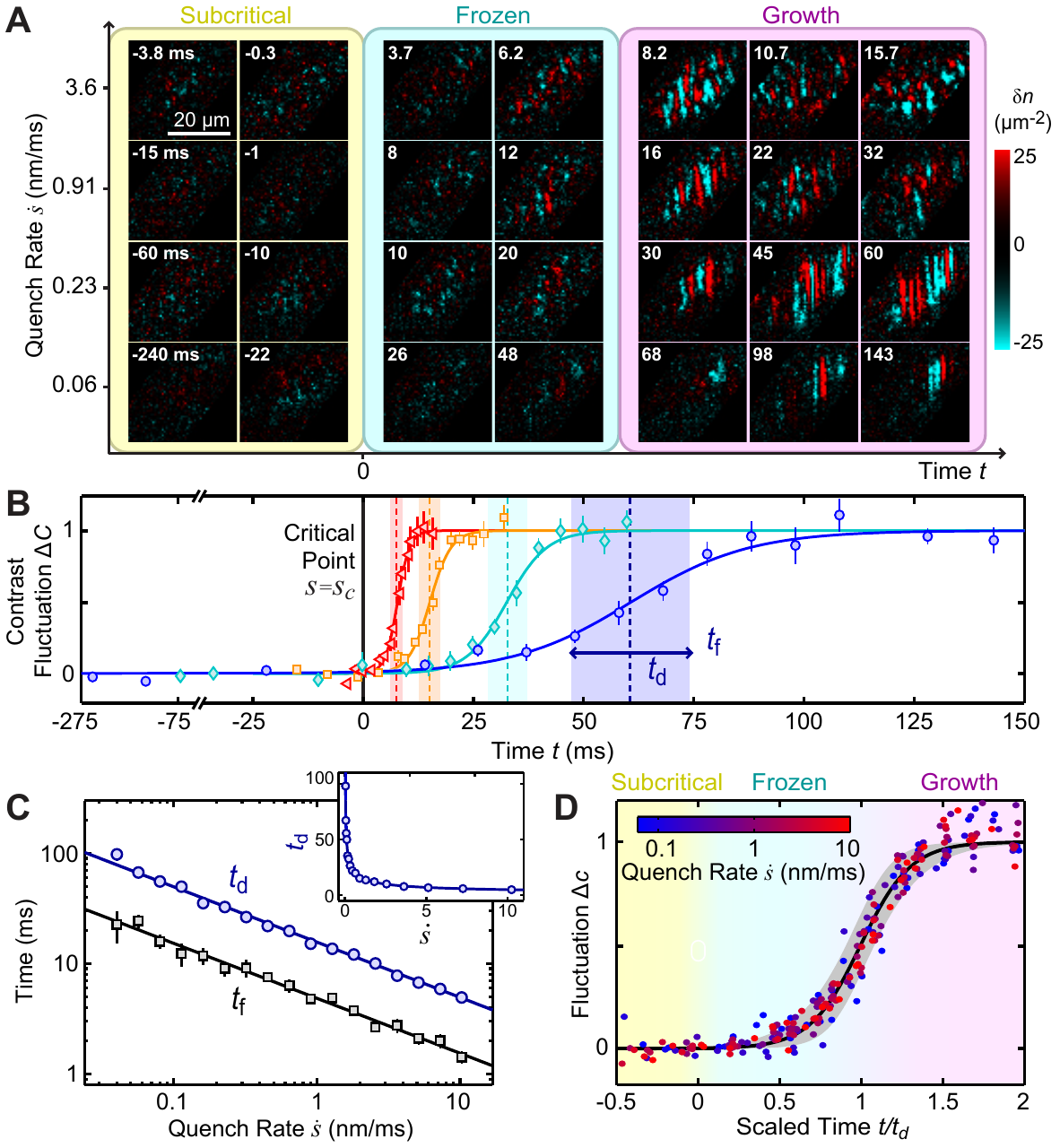}
\caption{\label{Fig2}Growth of quasimomentum fluctuation in quantum critical dynamics. (\textbf{A}) Sample images show the emergence of quasimomentum fluctuation throughout linear ramps across the ferromagnetic phase transition. Each ramp exhibits three regimes: a subcritical regime before the transition, a frozen regime beyond the critical point where the fluctuation remains low, and a growth regime in which fluctuation increases and saturates, indicating domain formation. Time $t=0$ corresponds to the moment when the system reaches the critical point. (\textbf{B}) Quasimomentum fluctuation for ramp rates $\dot{s}=$~3.6 (triangles), 0.91 (squares), 0.23 (diamonds), and 0.06 (circles) nm/ms arises at a delay time $t=t_\mathrm{d}$ over a formation time $t_\mathrm{f}$. Fluctuation is normalized for each ramp rate to aid comparison \cite{MaM}. The solid curves show fits based on Eq.~\ref{eqn:time_func}. (\textbf{C}) The dependence of $t_\mathrm{d}$ (circles) and $t_\mathrm{f}$ (squares) on the quench rate is well fit by power-laws (solid curves) with scaling exponents of $a_\mathrm{d}=0.50(2)$ and $a_\mathrm{f}=0.50(6)$, respectively. The inset shows $t_\mathrm{d}$ on a linear scale. (\textbf{D}) Fluctuations measured for 16 ramp rates from 0.06 to 10.3 nm/ms collapse to a single curve when time is scaled by $t_\mathrm{d}$ based on the power-law fit. The solid curve shows the best fit based on the empirical function (Eq.~\ref{eqn:time_func}), and the gray shaded region covers one standard deviation. Error bars in panels B and C indicate one standard error.}
\end{figure*}

The Kibble-Zurek mechanism provides a powerful insight into quantum critical dynamics. According to this theory, when the time remaining to reach the critical point inevitably becomes shorter than the relaxation time, the system becomes effectively frozen, see Fig.~\ref{Fig1}C. The system only unfreezes at a delay time $t_\mathrm{KZ}$ after passing the critical point, when relaxation becomes faster than the ramp. At this time topological defects form, and the typical distance $d_{\mathrm{KZ}}$ between neighboring defects is determined by the equilibrium correlation length. The Kibble-Zurek mechanism predicts that $t_{\mathrm{KZ}}$ and $d_{\mathrm{KZ}}$ depend on the quench rate $\dot{s}$ as

\begin{eqnarray}
t_{\mathrm{KZ}} \propto \dot{s}^{-a},~a = \frac{z \nu}{1 + z \nu},\label{eqn:time_scaling}\\
d_{\mathrm{KZ}} \propto \dot{s}^{-b},~b = \frac{\nu}{1 + z \nu},\label{eqn:space_scaling}
\end{eqnarray}

\noindent where $z$ and $\nu$ are the equilibrium dynamical and correlation length exponents given by the universality class of the phase transition. 

For slow ramps $t_{\mathrm{KZ}}$ and $d_{\mathrm{KZ}}$ diverge and become separated from other scales in the system, making them the dominant scales for characterizing the collective critical dynamics \cite{Kolodrubetz2012,Chandran2012,Francuz2016}. This idea motivates the universality hypothesis, which can be expressed as
\begin{equation}
\label{eqn:universal_observables}
f(x, t; \dot{s}) \propto F \left(\frac{x}{d_{\mathrm{KZ}}}~,~ \frac{t}{t_{\mathrm{KZ}}}\right),
\end{equation}
\noindent indicating that the critical dynamics of any collective observable $f$ obeys the scaling symmetry and can be described by a universal function $F$ of the scaled coordinates $x/d_{\mathrm{KZ}}$ and $t/t_{\mathrm{KZ}}$. The only effect of the quench rate is to modify the length and time scales. 

We test the scaling symmetry of time by monitoring the emergence of quasimomentum fluctuations at different quench rates. After loading the condensates into the lattice, we ramp the shaking amplitude linearly from $s=0$ to values well above the critical amplitude $s_c=13.1$~nm \cite{MaM} and interrupt the ramps at various times to perform a brief time-of-flight (TOF) before detection. After TOF the quasimomentum distribution of the sample can be extracted from the deviation $\delta{n}(\mathbf{r})$ in the density difference between the $+1$ and $-1$ Bragg diffraction peaks \cite{MaM}.
This detection method is particularly sensitive when the quasimomentum just starts deviating from zero, indicating the emergence of fluctuations in the ferromagnetic phase where the ground states have non-zero quasimomentum.

\begin{figure*}
\includegraphics[width=170mm]{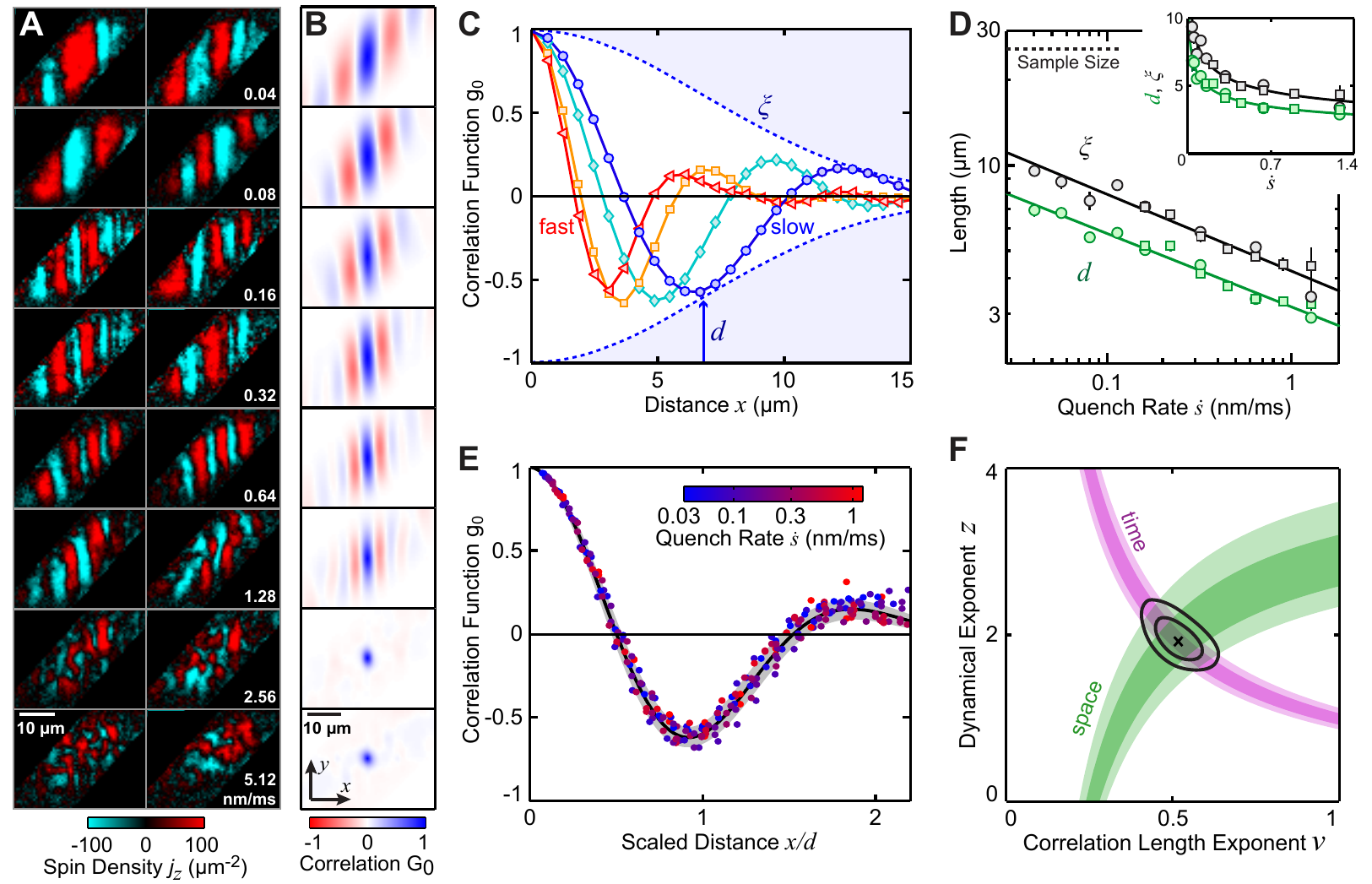}
\caption{\label{Fig3}Anti-ferromagnetic spatial correlations from quantum critical dynamics. (\textbf{A}) Sample images show spin domains measured near the time $t=1.4~t_\mathrm{d}$ after crossing the phase transition. (\textbf{B}) Spin correlation functions $G_0(\mathbf{r})=G(\mathbf{r})/G(0)$ (Eq.~\ref{eqn:spin_correlation}) are calculated from ensembles of 110-200 images. (\textbf{C}) Cuts across the density-weighted correlation functions $g_0(x)=g(x)/g(0)$ are shown for quench rates $\dot{s}=$~1.28~(triangles), 0.45~(squares), 0.16~(diamonds), and 0.056~nm/ms~(circles). Solid curves interpolate the data to guide the eye. The typical domain size $d$ and the correlation length $\xi$ are illustrated for 0.056~nm/ms by the arrow and dashed envelope, respectively. 
(\textbf{D}) The dependence of $d$ (green) and $\xi$ (black) on the quench rate is well fit by power-laws (Eq.~\ref{eqn:space_scaling}) with spatial scaling exponents of $b_\mathrm{d}=0.26(2)$ and $b_\mathrm{\xi}=0.26(5)$, respectively. We use two ramping protocols: linear ramps starting at $s=0$ (squares) and at $s=s_c$ (circles). The inset shows the results on a linear scale. Error bars indicate one standard error. (\textbf{E}) Correlation functions for $\dot{s}=0.04$-$1.28$~nm/ms collapse to a single curve when distance is scaled by the domain size extracted from the power-law fit. The solid curve shows the fit based on Eq.~\ref{eqn:space_func} while the gray shaded area covers one standard deviation. (\textbf{F}) The temporal scaling exponents $a_\mathrm{d}$ and $a_\mathrm{f}$ from Fig.~\ref{Fig2}C (magenta) and the spatial scaling exponents $b_\mathrm{d}$ and $b_\mathrm{\xi}$ from panel D (green) constrain the critical exponents $\nu$ and $z$ according to Eqs.~\ref{eqn:time_scaling}~and~\ref{eqn:space_scaling} with $68\%$~(dark) and $95\%$~(light) confidence intervals. The cross marks the best values with contours of $68\%$ and $95\%$ overall confidence. }
\end{figure*}

Over a wide range of quench rates the evolution of quasimomentum fluctuation can be described in three phases (Fig.~\ref{Fig2}A). First, below the critical point, quasimomentum fluctuation does not exceed its baseline level. Second, just after passing the critical point, critical slowing keeps the system ``frozen'', and fluctuation remains low. Finally, the system unfreezes and quasimomentum fluctuation quickly increases and saturates, indicating the emergence of ferromagnetic domains. We quantify this progression by investigating the fluctuation of contrast $\Delta{C}=\braket{\delta{n}^2/n^2}$ (Fig.~\ref{Fig2}B) that tracks quasimomentum fluctuation in our condensates \cite{MaM}, where $n$ is the total density and the angle brackets denote averaging over space and over multiple images.
We find empirically that the growth of fluctuations is well fit by the function
\begin{equation}
\label{eqn:time_func}
\Delta{C(t)}=\frac{1}{2}+\frac{1}{2}\textrm{tanh}\left(\frac{t-t_\mathrm{d}}{t_\mathrm{f}}\right),
\end{equation}
\noindent where the time $t$ is defined relative to when the system crosses the critical point at $t=0$, $t_\mathrm{d}$ characterizes the delay time when the system unfreezes, and $t_\mathrm{f}$ is the formation time over which the fluctuation grows. 

The measurement of fluctuation over time provides a critical test for both the Kibble-Zurek scaling and the universality hypothesis. First, both $t_d$ and $t_f$ exhibit clear power-law scaling with the quench rate $\dot{s}$ varied over more than two orders of magnitude (Fig.~\ref{Fig2}C). Power-law fits yield the exponents of $a_\mathrm{d}=0.50(2)$ and $a_\mathrm{f}=0.50(6)$, respectively. 
The nearly equal exponents are suggestive of the universality hypothesis, which requires all times to scale identically. Indeed, the growth of contrast fluctuation $\Delta{C}$ follows a universal curve when time is scaled by $t_\mathrm{d}$ (Fig.~\ref{Fig2}D), strongly supporting the universality hypothesis (Eq.~\ref{eqn:universal_observables}).

We next test the spatial scaling symmetry based on the structures of pseudo-spin domains that emerge after the system unfreezes. Here, we cross the critical point with two different protocols: the first is a linear ramp starting from $s=0$, while the second begins with a jump to $s=s_c$, followed by a linear ramp. We detect domains near the time $t=1.4~t_\mathrm{d}$ in the spin density distribution $j_z(\mathbf{r}) = n_+(\mathbf{r})-n_-(\mathbf{r})$ based on the density $n_{+/-}$ of atoms with spin up/down \cite{MaM}. At this time the spin domains are fully-formed and clearly separated by topological defects (domain walls), shown in Fig.~\ref{Fig3}A.
We characterize the domain distribution with the spin correlation function \cite{Sadler2006,Parker2013},
\begin{equation}
\label{eqn:spin_correlation}
G(\mathbf{r}) = \left<\int{j_z(\mathbf{R}+\mathbf{r}}) j_z(\mathbf{R}) d\mathbf{R}\right>, 
\end{equation}
\noindent averaged over multiple images, see Fig.~\ref{Fig3}B. Both ramping protocols lead to similar correlation functions, suggesting that the formation of topological defects does not depend on dynamics below the critical point.

The spin correlations reveal rich domain structure that strongly depends on the quench rate. For slower ramps $\dot{s}<1.3$~nm/ms the structures are predominantly one-dimensional and the density of topological defects increases with the quench rate. When the quench rate exceeds $1.3$~nm/ms, defects start appearing along the $y$-axis, and the domain structures become multi-dimensional. We attribute this dimensional crossover to the unfreezing time becoming too short to establish correlation in the non-lattice directions. For the remainder of this work we focus on the slower quenches and investigate the spin correlations along the lattice direction. 

We examine the one-dimensional correlations using line cuts of the density-weighted correlation functions $g(\mathbf{r})=G(\mathbf{r})/\braket{\int{n(\mathbf{R}+\mathbf{r}}) n(\mathbf{R}) d\mathbf{R}}$ \cite{Sadler2006,Parker2013}. The results exhibit prominent decaying oscillation, shown in Fig.~\ref{Fig3}C. We extract two essential length scales from the correlation functions: the average domain size $d$, or equivalently the distance between neighboring topological defects, and the correlation length $\xi$, indicating the width of the envelope function. These two scales are determined from the position and width of the peak in the Fourier transform of $g(x)$ \cite{MaM}.

These length scales enable us to test the spatial scaling symmetry. The lengths $d$ and $\xi$ both display power-law scaling consistent with the Kibble-Zurek mechanism, see Fig.~\ref{Fig3}D, with fits yielding exponents $b_\mathrm{d}=0.26(2)$ for the domain size and $b_\mathrm{\xi}=0.26(5)$ for the correlation length.
Furthermore, the correlations, measured at the same scaled time, collapse to a single curve in spatial coordinates scaled by the domain size $d$, see Fig.~\ref{Fig3}E. This result strongly supports the spatiotemporal scaling from the universality hypothesis (Eq.~\ref{eqn:universal_observables}). An empirical curve, 
\begin{equation}
\label{eqn:space_func}
g_0(x)=\exp{\left(-\frac{1}{2\sigma^2}\frac{x^2}{d^2}\right)}\mathrm{cos}\left(\frac{\pi}{\gamma}\frac{x}{d}\right),
\end{equation}
\noindent well fits the universal correlation function, yielding $\sigma$~=~1.01(1) and $\gamma$~=~1.04(1), indicating that the width of the envelope is close to the typical domain size. Furthermore, the Gaussian envelope can be characterized by a thermal length of $\lambda_\mathrm{s}=h/\sqrt{2\pi m k_\mathrm{B}T_\mathrm{s}}=\sqrt{4\pi}\sigma d$, where $k_\mathrm{B}$ is the Boltzmann constant and the effective spin temperature $T_\mathrm{s}\propto\dot{s}^{2b_\mathrm{d}}$ scales with the quench rate and reaches $20$~pK for our slowest ramps of $\dot{s}=0.04$~nm/ms.

The most striking feature of the universal correlation function is the emergence of oscillatory, anti-ferromagnetic order in the ferromagnetic phase.
In thermal equilibrium, ferromagnets are expected to have a finite correlation length but no anti-correlation.
We attribute the appearance of anti-ferromagnetic order in our system to the preferential generation of domains with a certain size during the quantum critical dynamics. A statistical analysis of the topological defect distribution reveals that the domain sizes are bunched with their standard deviation $\sigma_d=0.31(2)d$ well below their mean, indicating that the topological defects are created by a sub-Poisson process \cite{MaM}. 

Finally, the combined scaling exponents of space and time allow us to extract the equilibrium critical exponents based on the Kibble-Zurek mechanism, see Fig.~\ref{Fig3}F. Solving Eqs.~\ref{eqn:time_scaling} and \ref{eqn:space_scaling}, we obtain the dynamical exponent $z=$~1.9(2) and correlation length exponent $\nu=$~0.52(5), which agree with the mean-field values $z=2$ and $\nu=1/2$ within our experimental uncertainty. Note that the dynamical critical exponent $z=2$ results from the unique quartic dispersion $\epsilon=\beta{q_x^4}$ of our system at the critical point \cite{MaM}.

In summary, our experiment reveals a universal, spatiotemporal scaling symmetry of the dynamics across a quantum critical point. The observed scaling laws are in excellent agreement with the prediction from the Kibble-Zurek mechanism. Furthermore, the universal correlations exhibit intriguing anti-ferromagnetic order which would not be expected in equilibrium. Direct identification of the domain walls enables us to show that the anti-ferromagnetic correlations are connected to sub-Poisson generation of topological defects. The scaling of the correlation functions suggests that the anti-ferromagnetic order may be a shared feature of quantum critical dynamics for phase transitions in the same universality class, meriting future experiments. 

\begin{acknowledgments}
We thank L.-C. Ha, C. V. Parker, B. M. Anderson, B. J. DeSalvo, A. Polkovnikov, S. Sachdev, and Q. Zhou for helpful discussions. L.~W.~C. was supported by the NDSEG and Nambu Fellowships. This work was supported by NSF MRSEC (DMR-1420709), NSF Grant No.~PHY-1511696, and ARO-MURI Grant No.~W911NF-14-1-0003.
\end{acknowledgments}

\section{Supplementary Material}
\noindent\textbf{Experiment Setup.} Our condensates form in an optical dipole trap at the crossing of three lasers with wavelength $\lambda=1064$~nm. After evaporation the condensates are nearly pure, consisting of 30~000 to 40~000 atoms with temperatures less than $10$~nK. We adiabatically load the condensates into an optical lattice by retro-reflecting the trapping beam along the $x$-axis.
The apparatus which enables us to shake the lattice is described in Ref.~\textit{(28)}.

Our experiments rely on a careful choice of the parameters governing the shaking optical lattice. We set the shaking frequency $\omega=2\pi\times8.00$~kHz slightly above the zero-momentum band gap $h\times7.14$~kHz, such that shaking raises the energy near the center of the ground band.
During shaking we reduce the scattering length to $a=2.1$~nm using a Feshbach resonance to lower the heating rate \textit{(32)}. Finally, immediately before time-of-flight (TOF) we reduce the scattering length to $a=0$ to prevent collisions while the atoms separate into distinct Bragg peaks.

Based on the lattice depth and shaking frequency, we calculate the critical shaking amplitude $s_c=13.1$~nm using Floquet theory \textit{(28)}, above which the system acquires a double-well dispersion. We base our calculation of $a_d=0.50(2)$ on this theoretical critical point. To verify the critical value $s_c=13.1$~nm, we allow the critical shaking amplitude to be a free parameter in a power law fit of $t_\mathrm{d}$ while fixing the exponent to its theoretical value of 0.5. The fit yields $s_c=13.8(6)$~nm which is consistent with the calculated critical amplitude.

\subsection{Analysis of Quasimomentum Fluctuation}
\begin{figure*}
\begin{center}
\includegraphics[width=100mm]{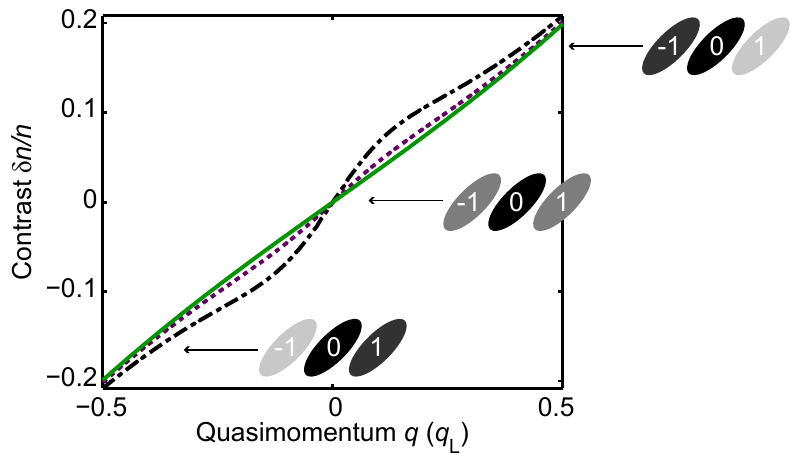}
\end{center}
\caption{\label{Fig4}Detection of quasimomentum via density deviation. The contrast $\delta{n}/n$ between the $\pm1$ Bragg peaks can be used to assess the quasimomentum $q$, according to a calculation of the Floquet eigenstates in the shaken lattice \textit{(28)}. The contrast only weakly depends on shaking amplitude, shown for $s=0$ (solid), $s=s_c$ (dotted), and $s=2s_c$ (dot-dashed). The illustrations represent the density in the three relevant Bragg peaks after time-of-flight. The lattice momentum is $q_\mathrm{L}=h/\lambda$ where $\lambda=1064$~nm is the wavelength of the lattice laser.}
\end{figure*}

We use the density fluctuation in the Bragg peaks to detect quasimomentum fluctuation in the gas. Non-zero local quasimomentum $q$ changes the local density difference between Bragg peaks. To detect this change we perform a brief TOF with duration $t_\mathrm{TOF}=5$~ms, which is long enough to separate the Bragg peaks but short enough that spatial information is preserved. From our images we calculate the density difference $\Delta{n}(\mathbf{r})=n_{-1}(\mathbf{r})-n_{1}(\mathbf{r})$ where $n_i(\mathbf{r})$ is the density of the $i$'th Bragg peak. To remove the small offset in $\Delta{n}$ which exists at momentum $q=0$, we calculate the density deviation $\delta{n}(\mathbf{r})=\Delta{n}(\mathbf{r})-\braket{\Delta{n}(\mathbf{r})}$, where the angle brackets denote averaging over multiple images. The shift $\delta{n}$ is nearly proportional to the local quasimomentum regardless of the shaking amplitude (Fig.~\ref{Fig4}). Finally, we calculate the contrast fluctuation $\Delta{C}=\braket{\delta{n}^2/n^2}$ which closely tracks quasimomentum fluctuation in our condensates, where $n(r)$ is the total density. The angle brackets denote averaging over many images and over the position within each sample.

In order to remove spurious sources of fluctuation such as photon and atom shot noise, we subtract the baseline value of $\Delta{C}$ for each ramp rate, which is given by the average of the three measurements at the earliest times taken below the critical point. Furthermore, even though quasimomentum fluctuation should continue to grow as $q^{*2}$ with increasing shaking amplitude, where $\pm{q^*}$ are the quasimomenta of the ground states, we find that $\Delta{C}$ appears to saturate to a nearly constant value for times well beyond $t_d$. We attribute saturation to the typical displacement $q^*t_\mathrm{TOF}/m$ during TOF becoming larger than the correlation length, such that fluctuation is dominated by the motion rather than the quasimomentum. We normalize $\Delta{C}$ to its saturated value at each ramp rate for convenient comparison. We determine the saturated value by the average of the latest three measured values, which are taken well beyond the delay time $t_d$.

\begin{figure*}
\begin{center}
\includegraphics[width=130mm]{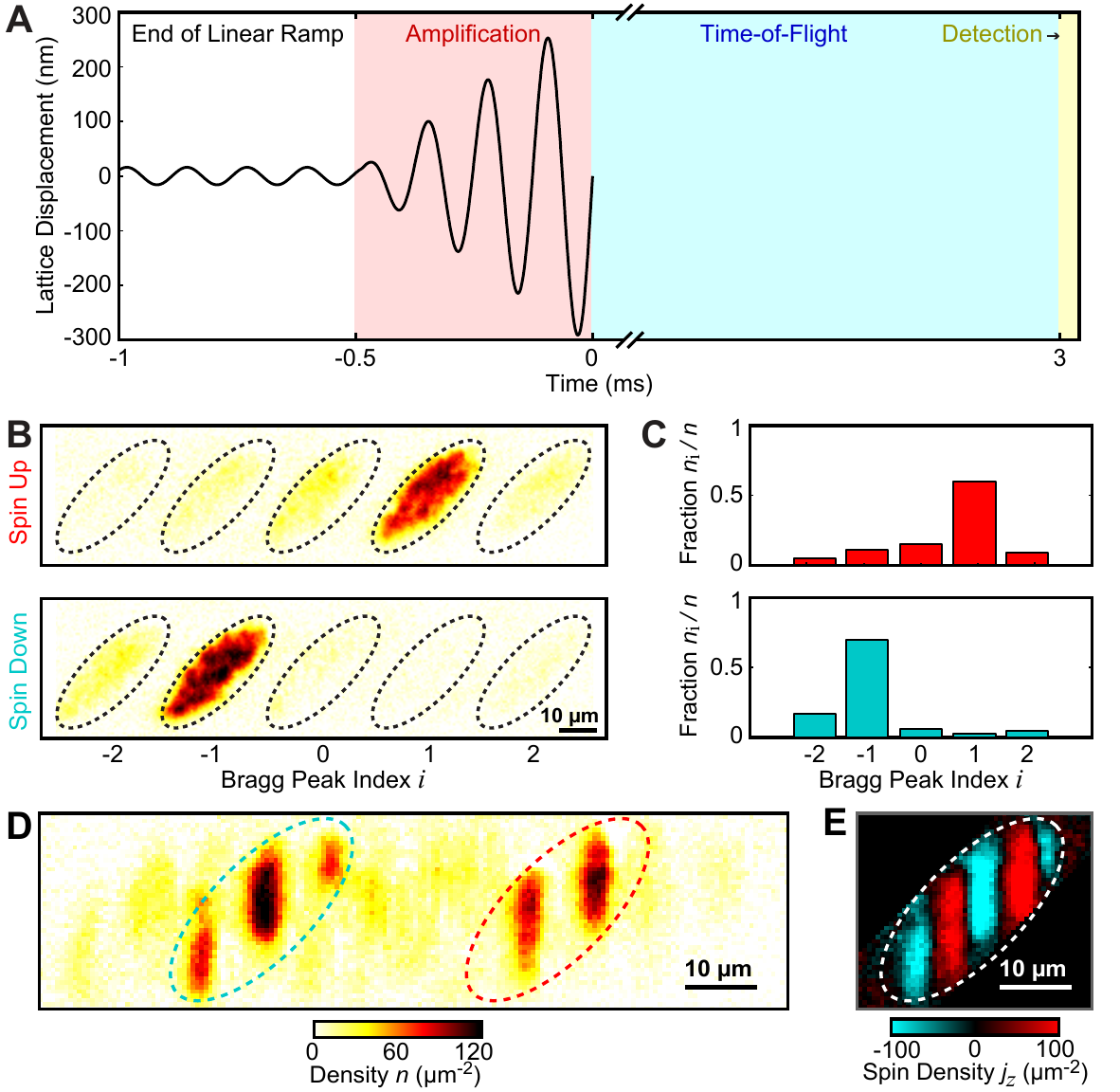}
\end{center}
\caption{\label{Fig5}Detection and reconstruction of ferromagnetic domains. (\textbf{A}) We amplify the distinction between the pseudo-spin states by rapidly increasing the shaking amplitude over $0.5~$ms before time-of-flight (TOF). After $3$~ms TOF we detect the density distribution by absorption imaging. The examples shown in this figure correspond to $s=32~$nm before amplification. (\textbf{B}) Sample images used to calibrate the occupation of each Bragg peak are taken with all of the atoms in pseudo-spin up (top) or down (bottom). Ellipses identify the Bragg peaks. For these images we use a longer TOF lasting $5$~ms. Each spin predominantly occupies a different Bragg peak. (\textbf{C}) The fraction of density $n_\mathrm{i}/n$ in each Bragg peak distinguishes spin up (top) and down (bottom). (\textbf{D}) A sample image shows the density distribution after $3$~ms TOF for a condensate with five domains. The ellipses mark the two most important Bragg peaks and are colored to indicate the spin state which dominates each peak. (\textbf{E}) Reconstruction based on the fractions in panel C produces the spin density distribution corresponding to the TOF image in panel D.}
\end{figure*}

\subsection{Reconstruction and Analysis of Pseudo-spin Domains}

\noindent\textbf{Reconstructing spin density.} Reconstruction enables us to study the \textit{in-situ} spin density distribution, including the domain structure. After linearly ramping the shaking amplitude at ramp rate $\dot{s}$ until $t=1.4~t_d$ when domains have fully formed, we rapidly increase the shaking amplitude by as much as a factor of 15 over only $0.5$~ms (Fig.~\ref{Fig5}A). The sudden increase in shaking amplifies the signal by exciting the two spin states to predominantly occupy different Bragg peaks. 
Depending on the shaking amplitude at $t=1.4~t_d$, we adjust the timing and the shaking amplitude immediately before release in order to maximize the distinguishability of the two spin states. We use images of condensates with uniform spin (Fig.~\ref{Fig5}B) to calibrate the projection of each spin state onto Bragg peaks (Fig.~\ref{Fig5}C). In comparison to the procedure without enhanced shaking used in Ref.~\textit{(28)}, the amplification stage improves the fraction of atoms which distinguish the spin states from 23\% to about 71\%, corresponding to an increase in signal by more than a factor of three. After the enhanced shaking period, we perform a $3~$ms TOF and measure the density in each Bragg peak (Fig.~\ref{Fig5}D), from which we can reconstruct the spin density distribution (Fig.~\ref{Fig5}E) using an algorithm similar to that described in Ref.~\textit{(28)}.\\

\noindent\textbf{Minimizing bias of the total polarization.} The effectively ferromagnetic quantum phase transition can be biased by a nonzero initial velocity of the condensate relative to the lattice \textit{(28)}. In order to focus our study on the dynamics across an unbiased quantum phase transition, we test the total spin polarization $P=\int{j_z(\vec{R})d\vec{R}}/\int{n(\vec{R})d\vec{R}}$ of each reconstructed image, which is expected to be close to zero for unbiased samples. Indeed, under most conditions ($0.16\leq\dot{s}<1.0 $~nm/ms) we find that more than $90\%$ of images have total polarization $|P|<0.3$. The correlation analysis excludes the remaining biased images with $|P|>0.3$. For very slow ramps ($\dot{s}<0.16~$nm/ms) starting from $s=0$, we find that many samples are biased, likely due to increased susceptibility to a small, uncontrolled velocity between the condensate and the lattice. We have excluded data from these conditions to avoid poor statistics.\\

\noindent\textbf{Removing systematic effects due to finite imaging resolution.}
We study the one-dimensional domain structures along the lattice direction ($x$) by taking cuts $g_{\mathrm{meas}}(x)$ along the long-axis of the measured, normalized correlation functions $g(\mathbf{r})$. Since the domain walls are predominantly oriented along the non-lattice direction ($y$), long axis cuts maximize the range of the correlation functions that we can evaluate but still reflect the structure along the lattice direction.

To obtain the physical spin correlations we must remove the systematic effects of our finite imaging resolution. Since the correlation functions depend on the spin density at both ends of the displacement vector, the measured correlations $g_{\mathrm{meas}}(x)$ are the physical correlations $g(x)$ convolved with the point spread function $P(x)$ twice \textit{(33)}.
We calculate the Fourier transform of the deconvolved correlation function $\tilde{g}(k)=\tilde{g}_{\mathrm{meas}}(k)/\tilde{P}^2(k)$ from the Fourier transforms of the measured correlations $\tilde{g}_{\mathrm{meas}}(k)$ and of the point spread function $\tilde{P}(k)$. Inverting the Fourier transform produces the correlation functions $g(x)$ shown in Fig.~\ref{Fig3}C. Furthermore, from the peak position $k_p$ in $\tilde{g}(k)$ we extract the typical domain size $d=\pi/k_p$, and from the full width at half maximum $\Delta{k}$ of the peak we extract the correlation length $\xi=\pi/\Delta{k}$.
\\

\begin{figure*}
\begin{center}
\includegraphics[width=130mm]{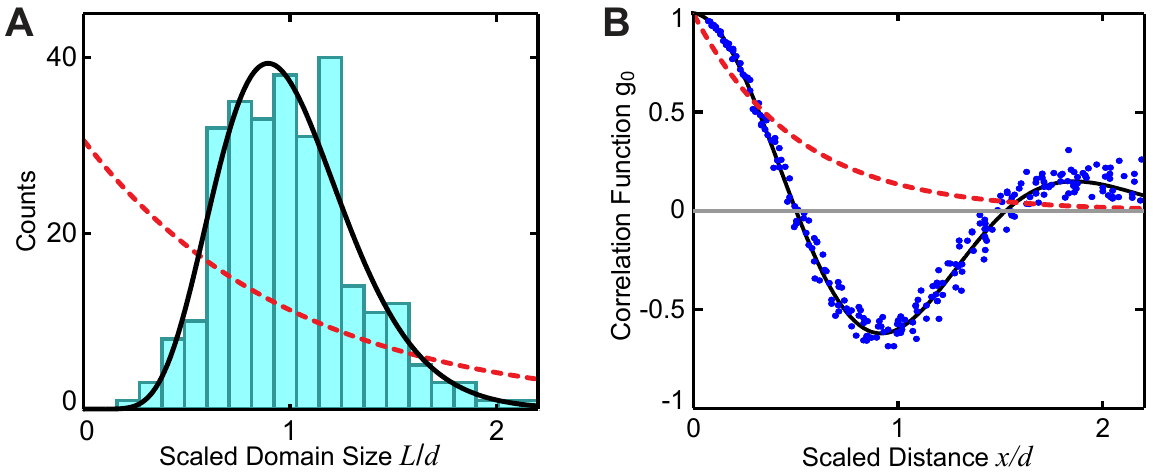}
\end{center}
\caption{\label{Fig6}Sub-Poisson generation of domain walls. (\textbf{A}) The distribution of domain sizes $L$ for 110 samples with quench rate $\dot{s}=0.08$~nm/ms is bunched near the average domain size $d$. The solid curve shows a fit based on the function $A (L/d)^{a-1} e^{-aL/d}$, where the coefficient $A=a^{a}/\Gamma(a)$, which interpolates between the exponential ($a=1$) and delta ($a\rightarrow\infty$) distributions. The fit yields $a=10(1)$ for the measured distribution. For comparison, the dashed curve shows an exponential distribution ($a=1$) corresponding to Poisson generation of defects. (\textbf{B}) Poisson generation of defects would lead to exponential decay of spin correlations as $g_\mathrm{P}(x)=e^{-2x/d}$ (dashed curve), which does not exhibit the anti-correlation seen in the data from Fig.~\ref{Fig3}E (blue points). The solid curve shows the fit to the measured correlations based on Eq.~7.}
\end{figure*}

\noindent\textbf{Sub-Poisson generation of domain walls} To better understand the process which generates domain walls we calculate the domain size distribution from our images. We identify domain walls by integrating the spin density along the y-direction, filtering noise at the single pixel scale (0.6~$\mu$m) which is below our resolution limit, and locating where the spin density changes sign. We calculate the domain sizes from the distances between neighboring walls.
Since the correlation functions in scaled space are invariant with quench rate we focus on a single rate $\dot{s}=0.08$~nm/ms, for which the domains are relatively large ($d=6.0~\mu$m) and easy to resolve. The resulting domain size distribution (Fig.~\ref{Fig6}A) is tightly bunched around its mean. This bunching would not be expected for a Poisson process, which should exhibit an exponential distribution due to the constant probability of forming a domain wall at any location. Similarly, Poisson generation of domain walls would lead to exponentially decaying correlations that are qualitatively distinct from the oscillatory correlations observed in our experiments (Fig.~\ref{Fig6}B).

\subsection{Equilibrium critical exponents $\nu$ and $z$}

The partition function $Z$ of our system near the critical point where $\alpha\rightarrow0$ and $\beta>0$ can be written in the path integral form as

\begin{eqnarray*}
Z&=&\int D\Psi D\Psi^* e^{-\int dxd\tau L} \\
L&=& \Psi^* \partial_{\tau} \Psi+\alpha |\partial_{x}\Psi|^2+\beta|\partial_{x}^2\Psi|^2-\mu\Psi^*\Psi+\frac{g}2(\Psi^*\Psi)^2,
\end{eqnarray*}

\noindent where $L$ is the mean field Lagrangian density, $\mu=g\rho_0$ is the chemical potential, $g$ is the interaction parameter and $\tau=i t/\hbar$.

Given a fluctuating order parameter $\Psi=\sqrt{\rho}e^{i\theta}$ around the equilibrium value $\Psi_0=\sqrt{\rho_0}$ and $\rho=\rho_0+\tilde{\rho}$, we have

\begin{align*}
\Psi^* \partial_{\tau} \Psi =& i \rho\partial_{\tau}\theta+\partial_{\tau}\rho/2\\
|\partial_{x}\Psi|^2 =&\rho(\partial_x\rho)^2+\frac1{4\rho}(\partial_x\theta)^2\\
|\partial_{x}^2\Psi|^2 =&\rho^{-3}\left(\frac{(\partial_x\rho)^2}4-\frac{\rho\partial_x^2\rho}2+\rho^2(\partial_x\theta)^2\right)^2\\
&~+\rho^{-1}(\rho \partial_x^2\theta-\partial_x\theta\partial_x\rho)^2\\
-\mu\Psi^*\Psi+\frac{g}2(\Psi^*\Psi)^2=&\frac g2(\tilde{\rho}^2-\rho_0^2).
\end{align*}

Eliminating terms like $\partial_{\tau}\rho$ and $\rho_0\partial_{\tau}\theta$ that contribute to constants after integration over $\tau$, we have

\begin{multline*}
L= i \tilde{\rho}\partial_{\tau}\theta+\frac{\alpha}{\rho}\left(\rho^2(\partial_x\rho)^2+\frac{(\partial_x\theta)^2}4\right) \\
+\frac{\beta}{\rho^3}\left(\frac{(\partial_x\tilde{\rho})^2}4  -\frac{\rho\partial_x^2\tilde{\rho}}2+\rho^2(\partial_x\theta)^2\right)^2\\
+\frac{\beta}{\rho}(\rho \partial_x^2\theta-\partial_x\theta\partial_x\tilde{\rho})^2+\frac g2\tilde{\rho}^2.
\end{multline*}

Since the amplitude excitations are gapped and the angular excitations are gapless, we can assume $\partial_x\tilde{\rho}=0$ in the long wavelength limit, which gives

\begin{eqnarray*}
L=i \tilde{\rho}\partial_{\tau}\theta+\alpha\rho\frac{(\partial_x\theta)^2}4+\beta\rho(\partial_x\theta^2)^2+\beta\rho(\partial_x^2\theta)^2+\frac g2\tilde{\rho}^2
\end{eqnarray*}

Completing the path integral over $\tilde{\rho}$, we obtain to leading order in $\theta$

\begin{eqnarray*}
L_\theta=\frac{1}{2g}(\partial_{\tau}\theta)^2+\frac{\alpha}{4}\rho_0(\partial_x\theta)^2+\beta\rho_0(\partial_x^2\theta)^2.
\end{eqnarray*}

The mean field correlation length exponent $\nu=1/2$ can be derived from the spatial scaling symmetry with $\partial_\tau\theta=0$:

\begin{eqnarray*}
\alpha\rightarrow\lambda^{-1}\alpha, x\rightarrow\lambda^{\nu} x.
\end{eqnarray*}

At the critical point $\alpha=0$, the dynamical critical exponent $z=2$ is determined from the following scaling symmetry:

\begin{eqnarray*}
x\rightarrow\lambda x, t\rightarrow \lambda^z t.
\end{eqnarray*}

\noindent Notably $z=2$ results from the dominance of quartic dispersion $\beta|\partial_{x}^2\Psi|^2$ at the critical point.

Given $z=2$ and $\nu=1/2$, the Kibble-Zurek temporal and spatial exponents from Eqs.~2 and 3 are given by $a=1/2$ and $b=1/4$, respectively. The results are in excellent agreement with our measured values of $a_{ex}=0.50(2)$ and $b_{ex}=0.26(2)$ within our experimental uncertainty.

\end{document}